\documentstyle[11pt,ystars_pasp,twoside]{article}
\begin{document}
\title{Can Post T Tauri Stars Be Found?  Yes!}
\author{Eric L. N. Jensen}
\affil{Swarthmore College, Department of Physics and Astronomy, 500
  College Ave., Swarthmore, PA 19081 USA}

\begin{abstract}
  I review the observational challenges of finding post T Tauri stars
  (PTTS), defined here as low-mass, pre--main-sequence stars with ages
  of $10^7$--$10^8$ yr.  Such stars are difficult to find because they
  are less active than younger T Tauri stars, and
  they may not be associated with molecular gas.  They are useful for
  studying the evolution of circumstellar disks and stellar activity
  between the $10^6$-yr ages of nearby star-forming regions and the
  main sequence.  However, care must be taken in the search process so
  that the selection criteria used to locate such stars do not bias
  the sample used for subsequent evolutionary studies.
\end{abstract}

\section{Introduction}

In an influential 1978 paper, Herbig raised the question ``Can Post T
Tauri Stars Be Found?''  (Herbig 1978).  More than twenty
years later, armed with the benefit of hindsight and data from
astrometric space missions and all-sky surveys at various wavelengths,
we can clearly answer ``yes'' to this question.  However, the problem
of finding such stars, particularly away from clusters, is still an
interesting and challenging one.  In this contribution, I discuss the
difficulties inherent in trying to identify pre--main-sequence stars
that are somewhat older than those found in regions of active
star formation and are located in the field.

\section{What is a Post T Tauri Star?}

Herbig (1978) noted that T Tauri stars have ``several distinct
observational characteristics,'' namely strong H$\alpha$ emission,
high surface lithium abundance, irregular variability, and excess
infrared emission, and he noted that these are ``diminished or
absent'' in low-mass main-sequence stars.  Thus, he defined PTTS as
those stars that display intermediate values of these characteristics.

The difficulty with adopting such a definition is that it begs the
question of how these properties evolve with time during the
pre--main-sequence evolution of a star.  By {\em assuming\/} evolution
as part of our definition, we are defining away our ability to study
those stars in which the evolution of one or more of the properties
listed above may be particularly fast or slow.  As a case in point, it
was not until the availability of abundant x-ray data from the
Einstein satellite in the 1980s that we realized that infrared excess
and strong H$\alpha$ emission are not ubiquitous features of T Tauri
stars (e.g., Walter 1986).  It took an observational
advance (in this case the launch of x-ray satellites) to allow
the detection of young stars based on a characteristic that
was not part of the original definition of the group, thus allowing
the study of the full range of infrared excesses exhibited by the T
Tauri population.

Thus, here I adopt an age-based definition of post-T-Tauri stars
(PTTS).  For the purpose of this review, I consider a PTTS to be a
low-mass pre--main-sequence star with an age of $10^{7}$--$10^{8}$ yr.
The lower bound corresponds roughly to the oldest stars found in
nearby active star-forming regions such as Taurus-Auriga and
Ophiuchus, while the upper bound is roughly the age at which
solar-mass stars reach the zero-age main sequence or ZAMS (e.g.,
Siess, Dufour, \& Forestini 2000).  If we use age as the
defining characteristic of a PTTS, then we are free to study the full
range of stellar and disk properties exhibited by such stars, and to
explore the evolution of these properties with stellar age.

The downside of using an age-based definition is that determining the
age of a star that is not in a cluster typically requires knowledge of
its distance in order to place it on a theoretical HR diagram for
comparison with evolutionary tracks.  However, the topic of this
meeting is young stars {\em near\/} Earth; we are
fortunate to live at a time when many of the nearby stars have
well-determined distances from Hipparcos, and many more will soon have
accurate distances from FAME (Greene, this volume) and GAIA\null.  Thus,
our prospects for determining the ages of a statistically significant
sample of nearby young stars are better than ever.

Unfortunately, the age of a star is not unambiguously determined by
its position in the HR diagram.  Stars may appear to lie above the
ZAMS for reasons that are both astrophysical and observational in
nature.  Most obviously, both pre--main-sequence and
post--main-sequence stars occupy this region of the HR diagram.  In
addition, observational uncertainties (including distance and
temperature errors, and the presence of unresolved binary companions)
can cause a main-sequence star to appear to lie above the ZAMS\null.
Finally, even the comparison of error-free observations with
theoretical tracks is not without difficulties.  Different sets of
tracks place the ZAMS at somewhat different positions in the HR
diagram (see, e.g., Stauffer, this volume), and transformation between
the observed quantities of apparent magnitude and spectral type (or
color) and the $L$ and $T_{\rm eff}$ of the theoretical HR diagram can
introduce uncertainties as well.

\section{Characteristics of PTTS}

Clearly, then, even when defining PTTS by age we need to understand
the unique observable characteristics of such stars in order to be
able to determine stellar ages and thus classify individual
stars unambiguously as PTTS\null.  In this section, I consider some
observable properties of PTTS that may help us distinguish them from
older stars.

At first glance, this appears to be almost exactly what I claimed
above we do {\em not\/} want to do: define a group of stars based on
one or more secondary characteristics, since this impairs our ability
to study the evolution of that characteristic.  However, I am not
abandoning an age-based definition; I am simply suggesting that, even
when using age as our defining characteristic, we must rely somewhat
on secondary characteristics to establish a young age observationally.
I will argue below that such an approach can still allow us to study
stellar and disk evolution if we are careful about choosing which
secondary characteristics to consider and if we are aware of how these
characteristics are interrelated.

Therefore, I now briefly review some distinguishing properties of
young stars, with a particular eye toward how such properties can help
us distinguish young from old stars, and how some of these properties
are interrelated.  Each of these properties has been the subject of
numerous reviews in the literature.  The aim here is not to
cover each property exhaustively, but merely to provide an overview of
how it fits into the study of PTTS\null.  Variability and kinematics
are also useful indicators of youth, but they are not discussed below
for reasons of limited space.

Lest it get lost in the details that follow, I stress the following
basic point. Position in the HR diagram is most useful for
distinguishing between pre--main-sequence and ZAMS stars; secondary
indicators are less useful for this.  On the other hand, secondary
indicators such as lithium and x-ray emission are most useful for
distinguishing between pre-- and post--main-sequence stars; position
in the HR diagram is less useful for this.  Thus, the combination of
HR diagram position and one or more of the following secondary
indicators of youth yields the most effective strategy for finding
PTTS\null.

\subsection{Infrared excess}

At ages of $10^{6}$ yr, more than 50\% of low-mass stars show excess
infrared emission above their photospheric emission (Meyer \& Beckwith
2000; Haisch, Lada, \& Lada 2001).  This
emission arises in circumstellar disks.  At PTTS ages, the fraction of
stars retaining substantial disks is very uncertain; this question is
the subject of active study (e.g., Spangler et al.\
2001; Meyer \& Beckwith 2000).  Certainly some
PTTS still show evidence for disks, with two notable examples being TW
Hya (Rucinski \& Krautter 1983) and HD 98800 (Zuckerman \&
Becklin 1993).  At the $\sim 10^{8}$ yr age of the Pleiades,
very few stars have detectable infrared excess (Meyer \& Beckwith
2001).  While a small fraction ($< 1$\%) of giants have infrared
excesses (de la Reza, Drake, \& da Silva, 1996), these stars
are rare, so the presence of a strong infrared excess is in general an
indicator of youth.

\subsection{H$\alpha$ emission}

Strong H$\alpha$ emission is another defining characteristic of the
youngest stars.  T Tauri stars show a range of emission-line
strengths, with H$\alpha$ emission equivalent widths ranging from a
few \AA ngstroms up to several hundred \AA ngstroms.  The stars with
the strongest H$\alpha$ emission lines (EW $\mathrel{\hbox{\rlap
    {\hbox{\lower4pt\hbox{$\sim$}}}\hbox{$>$}}}$ 10 \AA) almost
invariably show infrared excesses as well, and it is believed that the
emission arises from the heating of disk material as it is accreted
onto the star.  Weaker emission lines (or filled-in absorption lines)
can arise from chromospheric activity.  Strong H$\alpha$ emission is
clearly tied with the presence of disks in PTTS as well; it was the
strong H$\alpha$ emission of TW Hya in the objective prism survey of
Henize (1976), coupled with its high galactic latitude, that
first brought it to the attention of Herbig, causing him to label TW
Hya as a candidate PTTS (Herbig, personal communication).

\subsection{X-ray emission}

Young stars as a group have a higher mean x-ray luminosity than older
stars; comparison of clusters of different ages shows that the median
x-ray luminosity declines steadily with age from $10^{6}$ yr to at
least $10^{8}$ yr (Brice\~no et al.\ 1997).  The
distance-independent ratio $L_{\rm X}/L_{\rm bol}$ also declines with
age.

The spread of x-ray luminosities at a given age is such that it is
difficult to distinguish ZAMS stars from pre--main-sequence stars on
the basis of x-ray luminosity or $L_{\rm X}/L_{\rm bol}$ alone
(Brice\~no et al.\ 1997).  However, giants with strong x-ray
emission are very rare, so a high $L_{\rm X}/L_{\rm bol}$ ratio is a
useful discriminant between pre-- and post--main-sequence stars.  The
exception to this is that short-period binaries maintain high levels
of activity throughout their lives due to the tidal locking of the
stellar rotation periods with the binary orbital period.

\subsection{Lithium abundance}

Lithium is a tracer of stellar youth since it is destroyed at
temperatures of around $10^{6}$ K, and late-type stars have outer
convection zones that carry surface material down to layers of the
star with these temperatures.  As convection carries the
lithium-depleted material back to the surface, the photospheric
lithium abundance in these stars steadily declines with age.  Thus,
the presence of a strong lithium line in the spectrum of a star with a
spectral type of roughly G5 or later is a good indicator of youth. 

However, there are a few caveats to this statement.  First, stars with
spectral types earlier than about G0 do not have deep enough
convection zones to deplete lithium significantly, and G0--G5 stars
deplete lithium very slowly.  Also, at a given age, there is a
range of observed lithium equivalent widths at any given spectral type
(e.g., Soderblom et al.\ 1993).  This seems to be tied to
rotation rate, with rapidly rotating stars depleting lithium less
quickly.  Jeffries (1999) has suggested that in fact there may
be no spread in lithium {\em abundance\/} at a given spectral type and
age, and that the observed Li equivalent width spread is instead
accounted for by a spread in the fraction of a star covered by spots.
There is also a small population of lithium-rich giant stars; these
comprise $\mathrel{\hbox{\rlap
{\hbox{\lower4pt\hbox{$\sim$}}}\hbox{$<$}}} 1$\% of giants (Brown et
al.\ 1989).  There is some question in the literature about
whether or not  the Li-rich giants tend to be rapid
rotators (De Medeiros et al.\ 2000; Charbonnel \&
Balachandran 2000).  Some of the Li-rich giants
are also those that show infrared excesses (de la Reza et al.\ 1996).

\section{PTTS Search Strategies}

Given these properties, how then should one search for PTTS? 
There is no single answer to this question.  Given the rich set of
PTTS properties discussed above, clearly there are many ways that one
could proceed, with the optimum search strategy depending on the
astrophysical questions that one wishes to use these stars to
study.

The conclusions of Wichmann (2000) are of great relevance for
any survey for isolated PTTS\null.  He points out that any survey for
objects that are intrinsically rare (including PTTS, since the
pre--main-sequence phase is a small fraction of a star's life) is much
more likely to yield false positives (old stars misidentified as
young) than it is to yield false negatives (young stars misidentified
as old).  However, he also notes that the frequency of false PTTS
positives declines dramatically when selection based on both lithium
and x-ray emission is used.  More generally, that is one of the 
fundamental points of this paper.  Given the complicated nature of 
the PTTS and the degeneracy of many of the secondary selection 
criteria, use of multiple criteria is not only useful, it is nearly 
essential in building a convincing case that a star is a PTTS\null.

\subsection{Studies of disk evolution}

One of the exciting uses of a sample of nearby PTTS would be to allow
detailed study of disk properties in the planet-building phase at ages
of $10^7$--$10^8$ yr.  Some early surveys for PTTS, such as the Pico
dos Dias survey (de la Reza et al.\ 1989; Gregorio-Hetem et
al.\ 1992; Torres et al.\ 1995), used IRAS fluxes as
a selection criterion.  Thus, these surveys are biased toward stars
(such as TW Hya) with long-lived disks.  In order to study the
evolution of disks, we must select stars in a way that is unbiased
with respect to disks.  From the criteria above, perhaps the best
candidates are x-ray emission and Li abundance, neither of which
should be greatly influenced by the presence of disks.

In an attempt to establish a ``clean'' sample of PTTS for studying
disk evolution, I have begun a survey of stars selected for high x-ray
activity and position above the main sequence using the ROSAT Bright
Source Catalog and the Hipparcos Catalog (Jensen et al., in
preparation).  This approach is limited by the magnitude limit of
Hipparcos, but it holds great promise for use with FAME\null.  This
x-ray and optical selection recovers a number of known PTTS (including
members of the TW Hya and Tucana associations), suggesting that it is
effective for finding PTTS\null.  Follow-up spectroscopy of stars not
already known to be young has revealed a handful of stars with strong
Li absorption and H$\alpha$ emission.  Preliminary analysis suggests
that most of these are loosely associated with the large Sco-Cen-Lupus
molecular cloud complex or the Carina-Vela moving group (Makarov \&
Urban 2000).

\subsection{Studies of multiplicity or stellar rotation}

In contrast to studying disk evolution, studying the evolution of
multiplicity (at least at the shortest periods) or stellar rotation
with a sample of field PTTS is problematic.  Stellar rotation
is tied to multiplicity in the sense that short-period spectroscopic
binaries become tidally locked, causing the stars to rotate with the
binary orbital period.  This high rotation rate in turn leads to
stronger-than-average chromospheric and coronal activity (yielding a
high x-ray luminosity) and may delay the depletion of lithium.  Thus,
any sample that is selected based on lithium and/or x-ray criteria may
be biased toward having a high fraction of spectroscopic binaries and
rapidly-rotating single stars.  On the other hand, these concerns may
not apply to multiplicity at wider separations, unless the presence of
a short-period companion reduces the likelihood that a star also has a
wider companion.

\section{Conclusions}

The essential point of this contribution is that post T Tauri stars
{\em can\/} be found, but that their properties are complex and
interrelated.  Few if any of these properties uniquely determine the
age of a star, so secure classification of any field star as post T
Tauri must rest on the use of multiple criteria, most powerfully a
combination of position in the HR diagram and one or more of the
secondary criteria discussed above.  In choosing which criteria to use
in PTTS searches, we must carefully consider how these criteria relate
to (and possibly bias) the astrophysical properties we wish to study
using the PTTS.

\acknowledgments

I thank George Herbig for inspiring me to enter this field of study 
with his 1978 paper, and I apologize for borrowing that paper's title 
and pressing it into service for my own purposes.  I also thank him 
for answering my questions about his early studies of TW Hya.  I thank
Keivan Stassun, Rabi Whitaker, and David Cohen for useful conversations.

\end{document}